\documentclass[aps,onecolumn,preprint,superscriptaddress,showpacs,nofootinbib]{revtex4}
\pdfoutput=1
\usepackage{amsmath,amssymb,graphicx} 
\usepackage{epsf}
\usepackage{pstricks}

\newcommand{\beq}{\begin{eqnarray}}
\newcommand{\eeq}{\end{eqnarray}}

\newcommand{\gsim}{ \mathop{}_{\textstyle \sim}^{\textstyle >} }
\newcommand{\lsim}{ \mathop{}_{\textstyle \sim}^{\textstyle <}}

\newcommand{\BR}{\mbox{BR}}
\begin{document}

\title{Searching for the light dark gauge boson in GeV-scale experiments}

\author{Matthew Reece}
\email{mreece@princeton.edu}
\affiliation{Princeton Center for Theoretical Science, Princeton University, Princeton, NJ, 08544}

\author{Lian-Tao Wang}
\email{lianwang@princeton.edu}
\affiliation{Department of Physics, Princeton University, Princeton, NJ, 08540}

\date{\today}

\begin{abstract}
We study current constraints and search prospects for a GeV scale vector boson at a range of low energy experiments. It couples to the Standard Model charged particles with a strength $\leq 10^{-3} - 10^{-4}$ of that of the photon. The possibility of such a particle mediating dark matter self-interactions has received much attention recently. We consider searches at low energy high luminosity colliders, meson decays, and fixed target experiments. Based on available data, searches both at colliders and in meson decays can discover or exclude such a scenario if the coupling strength is on the larger side. We emphasize that a dedicated fixed target experiment has a much better potential in searching for such a gauge boson, and outline the desired properties of such an experiment. Two different optimal designs should be implemented to cover the range of coupling strength $10^{-3}-10^{-5}$,  and $<10^{-5}$ of the photon, respectively.  We also briefly comment on other possible ways of searching for such a gauge boson. 
\end{abstract}

\maketitle

\section{Introduction}
\label{sec:intro}
\setcounter{equation}{0}
\setcounter{footnote}{0}

The possibility that there is a light gauge boson that couples to the SM charged particles with a much suppressed coupling has been considered in various contexts 
\cite{Holdom,Ubosonrefs,gninenko,gninenko-g2,fayet,dorokhov,Redondo:2008ec,zhu-bes,Borodatchenkova:2005ct,HyperCPSM,UbosonHyperCP,PospelovSecluded}. It has been dubbed the U-boson. 
Recent data from PAMELA \cite{pamela}, ATIC \cite{atic}, and PPB-BETS \cite{ppb-bets} have motivated considering this possibility in a new light \cite{ArkaniHamed:2008qp,Pospelov:2008jd}. Similar scenario has been considered earlier in a different context \cite{Pospelov:2007mp}. Further modeling has appeared in Ref. \cite{Chun:2008by,models,Katz:2009qq}. See also \cite{Meade:2009rb,Batell:2009yf,Essig:2009nc}. We will look at how it can be probed in various experiments.

There are several ways that such a weakly interacting particle can arise \cite{ArkaniHamed:2008qp}. In this article, we consider the minimal setup in which the U-boson is the gauge boson of an additional ``dark'' $U(1)_d$ which has small kinetic mixing with the Standard Model $U(1)_{\rm Y}$. Such kinetic mixing can naturally arise if there are states charged under both $U(1)$s, even if such states are very heavy. This is the lowest order such coupling we can write which is consistent with all known symmetries of the Standard Model. 

We survey mainly three classes of experiments: low energy ($E_{\rm cm} \lsim 10$ GeV) colliders, rare decay channels at meson factories, and fixed target experiments. We find that the searches at both the low energy colliders and meson factories are capable of probing the parameter region of interests if the coupling size is on the larger side. We collect the useful search channels. At the same time, we find that carefully designed fixed target experiments, with the potential of large enhancement of luminosity, offer the chance to completely cover the interesting parameter region for a U-boson heavy enough to decay to two muons. We outline the necessary properties of such a setup in section~\ref{sec:fixed}.
\section{U-boson coupling}

We consider an additional $U(1)_d$ gauge boson, $a_{\mu}$, which is dubbed the U-boson. Its coupling to the Standard Model is induced by gauge kinetic mixing with $U(1)_{\rm Y}$ of the SM. After electroweak symmetry breaking, $U(1)_d$ mixes with the SM photon through
\beq
\mathcal{L}_{\rm kin-mix} = -2\epsilon F_d^{\mu \nu} F_{\mu \nu}.
\eeq
We can eliminate this mixing by the field redefinition $A_{\mu} \rightarrow A_{\mu} + \epsilon a_{\mu}$. At the same time, an interaction with the SM charged fermions
\beq
\epsilon a_{\mu} J^{\mu}_{\rm EM}
\eeq
will be generated, where $J^{\mu}_{\rm EM}$ is the Standard Model electromagnetic current. $U(1)_d$ is assumed to be broken spontaneously so that the mass of the U-boson $m_{\rm U}\sim $ GeV. This phase transition is achieved by introducing a dark higgs $h_d$ which acquires a GeV scale vev. 

The U-boson also couples to  the SM weak neutral current, with a strength further suppressed by a factor of $m_{U}^2/m_{\rm Z}^2$. At the same time, we have an order $\epsilon$ coupling between the SM Z-boson and the $U(1)_d$ current $J_{d}$, 
\beq 
- \epsilon Z_{\mu } \sin \theta_W J_d^{\mu}.
\eeq
Since $U(1)_d$ is assumed to be spontaneously broken at the GeV scale, $J_d$ minimally contains the dark higgs. 

Many closely related, but distinct, scenarios have received some attention in the literature. The U-boson we consider has very small axial couplings to SM matter, so many signatures discussed in Refs. \cite{fayet, dorokhov} are not expected to apply here. Furthermore, we do not expect it to decay invisibly with high branching fraction, although this depends on details of the hidden sector. Phenomenology involving such invisible decays has been discussed in Ref. \cite{invisible}. Scenarios where the U-boson couples in the way that we have described, but its mass is several orders of magnitude lighter, have distinct phenomenology as well \cite{subeV}.

We will consider the case $\epsilon \sim 10^{-4} - 10^{-3}$. This is the natural size of this coupling if we assume it originates from loops of heavy particles. 
Parametrically, this is $\sim \frac{g_Y g_d}{16\pi^2} \log\frac{M^2}{M'^2}$, with $M, M'$ mass scales of heavy particles coupling to the two gauge groups. While $\epsilon \sim 10^{-2}$ is possible if the log is large, one generally expects the scales $M$ and $M'$ to not be too separated. (For instance, in a GUT model, $M$ and $M'$ would be masses of different members of the same GUT multiplet.  These would both be near the GUT scale, not GUT and TeV respectively.)  

It is also roughly the interesting range for this scenario to explain the DAMA signal \cite{dama} when the U-boson is embedded in a scenario with inelastic dark matter \cite{idm,ArkaniHamed:2008qp}, which determines the quantity $\alpha_D \epsilon^2$ where $\alpha_D$ is the coupling in the dark sector. The range that explains DAMA, for $\alpha_D \approx \alpha$, is roughly $\epsilon \sim 10^{-4} \left(\frac{m_U}{1~{\rm GeV}}\right)^2$. A more detailed discussion appears in Ref.~\cite{Essig:2009nc}.

\section{Reach of U-boson searches at low energy high luminosity colliders}
\label{sec:babar}
\setcounter{footnote}{0}

In this section, we examine the constraints and discovery potential for the U-boson at GeV-scale collider experiments. The U-boson couples dominantly to $J_{\rm EM}$. Its production at colliders is identical to that of the photon, although with a much suppressed rate. Therefore, any experiment which produces a large number of energetic photons will have a chance to produce and detect U-bosons as well. 

Before a detailed numerical study,  we present an order of magnitude estimate of the reach at such low energy colliders (including BaBar, Belle, and BEPC). The U-boson signal comes from the production  process $e^+ e^- \rightarrow \gamma U$,  followed by $U \rightarrow \ell^+ \ell^-$. We denote the rate of this signal process $\sigma_{\rm s}$. The analogous QED process $e^+ e^- \rightarrow \gamma \gamma$ has a rate $\sigma_0 \sim \sigma_{\rm s} / \epsilon^2$.  The main background in this case is QED process $e^+ e^-  \rightarrow \gamma \ell^+ \ell^{-}$, with the invariant mass of the lepton pair $m_{\ell^+ \ell^-}\sim m_U$. The dominant process is $e^+ e^-  \rightarrow \gamma \gamma^* \rightarrow \gamma \ell^+ \ell^-$, with $q_{\gamma^*} \sim m_U$. The total rate for the QED background can be estimated as $\sigma_{3} \sim (\alpha /\pi) \sigma_0 \log (E_{\rm CM} / 2 m_{\ell})$. The background rate in a window of size $\delta m$ around $m_{\ell^+ \ell^-} = m_U$ is 
\beq
\Delta \sigma_3 \sim \frac{\alpha}{\pi} \sigma_0 \frac{\delta m}{m_{U}} =
\frac{\alpha}{\pi} \frac{\sigma_{\rm s}}{\epsilon^2} \frac{\delta m}{m_{U}}.
\eeq
Therefore, with integrated luminosity $\mathcal{L}$, we have
\beq
\label{eq:collider-reach}
\frac{\mbox{S}}{\sqrt{\mbox{B}}} \sim \sqrt{ \sigma_0 \mathcal{L}} \frac{\epsilon^2}{\sqrt{\alpha / \pi}} \sqrt{\frac{m_{U}}{\delta m}} \times \mbox{ BR}(U \rightarrow \ell^+ \ell^-).
\eeq
$\sigma_0 \sim 1 \times 10^7 $fb. We conclude that with 1 ab$^{-1}$ of integrated luminosity, we can at most achieve sensitivity $\epsilon \leq 10^{-4}$  with an idealized detector ($\delta m \approx 1$ MeV and perfect particle identification efficiency) for a $U$ boson that decays dominantly into leptons. In reality, the decay branching ratio into hadronic states can be extracted from the R-value measured in $e^+ e^- \rightarrow$ hadrons processes \cite{Batell:2009yf}. There is also another production channel $e^+ e^- \to e^+ e^- (U \to \ell^+ \ell^-)$, and the corresponding QED background $e^+ e^- \to e^+ e^- \ell^+ \ell^-$. Although higher order in coupling, the rate of these processes are non-negligible due to large forward and collinear enhancements. However, as we will demonstrate in our careful analysis in the next section, such channels turn out to give minor enhancement  of the reach at low $m_U$ and can be ignored for larger $m_U$. 

We emphasize that, from Eq.~\ref{eq:collider-reach}, the reach on $\epsilon$ improves by a factor $\mathcal{L}^{1/4}$ with increased luminosity. Therefore, accumulation of statistics can only  extend the reach for the U-boson at colliders slowly.

We pause here to comment on earlier low energy colliders. Many low energy resonances have been discovered this way.  Unfortunately, none of them has enough luminosity to discover the U-boson with such weak couplings. Previous low and medium energy lepton colliders, such as DCI, SPEAR, VEPP 4, DORIS, PEP, PETRA and TRISTAN, with center of mass energy ranging from several to several tens of GeV, typically have luminosity of $10^{30}-10^{31} $ cm$^{-2}$s$^{-1}$, or $10-100$ pb$^{-1}$yr$^{-1}$. Therefore, following our estimate in Eq.~\ref{eq:collider-reach}, they are insufficient to search or set an interesting bound. A similar conclusion applies to $pp$ colliders such as ISR. We can also estimate the production potential of the U-boson by scaling from the data of known profusely produced mesons. From the best measured $\rho$ decay branching ratios, $\BR (\rho \to \mu^+ \mu^-) = (4.55 \pm 0.28) \times 10^{-5}$, we can estimate about $10^7$ $\rho$ mesons have been produced. However, the on-shell production rate of the U-boson is much smaller, by $\epsilon^2 \times \Gamma_U / \delta E_{\rm cm}$, where beam energy spread $\delta E_{\rm cm}$ is typically 1-10 MeV. Due to its small coupling to the SM fields, the U-boson is typically very narrow, $\Gamma_{U} \leq 10^{-2} $ keV. Therefore, previous low energy collider searches do not have enough integrated luminosity to   produce the U-boson.

Sometimes, the interference between a resonance and continuum background can provide useful search channels, as it is linearly proportional to the coupling constant. However, such a signal is not available in our case. The resonance under our consideration is very narrow: $\Gamma_U$ is much smaller than the typical detector resolution, $\delta M \geq $ MeV. Since the main interference effect only exists within several widths around the resonance, such an effect is completely washed out by the detector resolution. 

\subsection{Search at BaBar}
\label{sec:babar-reach}
The reach at $B$-factories for a light U-boson has been previously estimated \cite{Borodatchenkova:2005ct}; we carry out a more careful study  and understand the mass dependence, including the processes $e^+ e^- \to U\gamma$ with $U \to e^+ e^-,~\mu^+ \mu^-,~{\rm or}~\pi^+ \pi^-$. For concreteness, we'll begin by discussing BaBar.

BaBar collides a 9.0 GeV electron beam on a 3.1 GeV positron beam. We have simulated the $\ell^+ \ell^- \gamma$ backgrounds in MadGraph \cite{MadGraph}. To search for a narrow resonance, it is important to understand the mass resolution of the detector. Using detector resolutions from  Ref.\cite{BaBarDetector}, we obtain the following empirical relations:
\beq
\delta m(e^+e^-) & = & \left(2.0 + 3.9 \left(\frac{m_U}{1.0~{\rm GeV}}\right) + 0.25 \left(\frac{m_U}{1.0~{\rm GeV}}\right)^2 \right) {\rm MeV}, \\
\delta m(\mu^+\mu^-) & = & \left(1.8 + 4.1 \left(\frac{m_U}{1.0~{\rm GeV}}\right) + 0.28 \left(\frac{m_U}{1.0~{\rm GeV}}\right)^2 \right) {\rm MeV}, \\
\delta m(\pi^+ \pi^-) &=& \left(0.65 + 5.3 \left(\frac{m_U}{1.0~{\rm GeV}}\right)\right){\rm MeV},
\eeq
where the last is restricted to a small region near the $\rho$ mass. See appendix~\ref{app:resolution} for a detailed discussion of how these numbers were obtained. The only cuts that we have applied to the background are basic geometric acceptance cuts for tracks and photons, and the requirement that track $p_t \geq 60$ MeV and $E_\gamma \geq 20$ MeV.

\begin{figure}[!h]
\begin{center}
\includegraphics[scale=0.27]{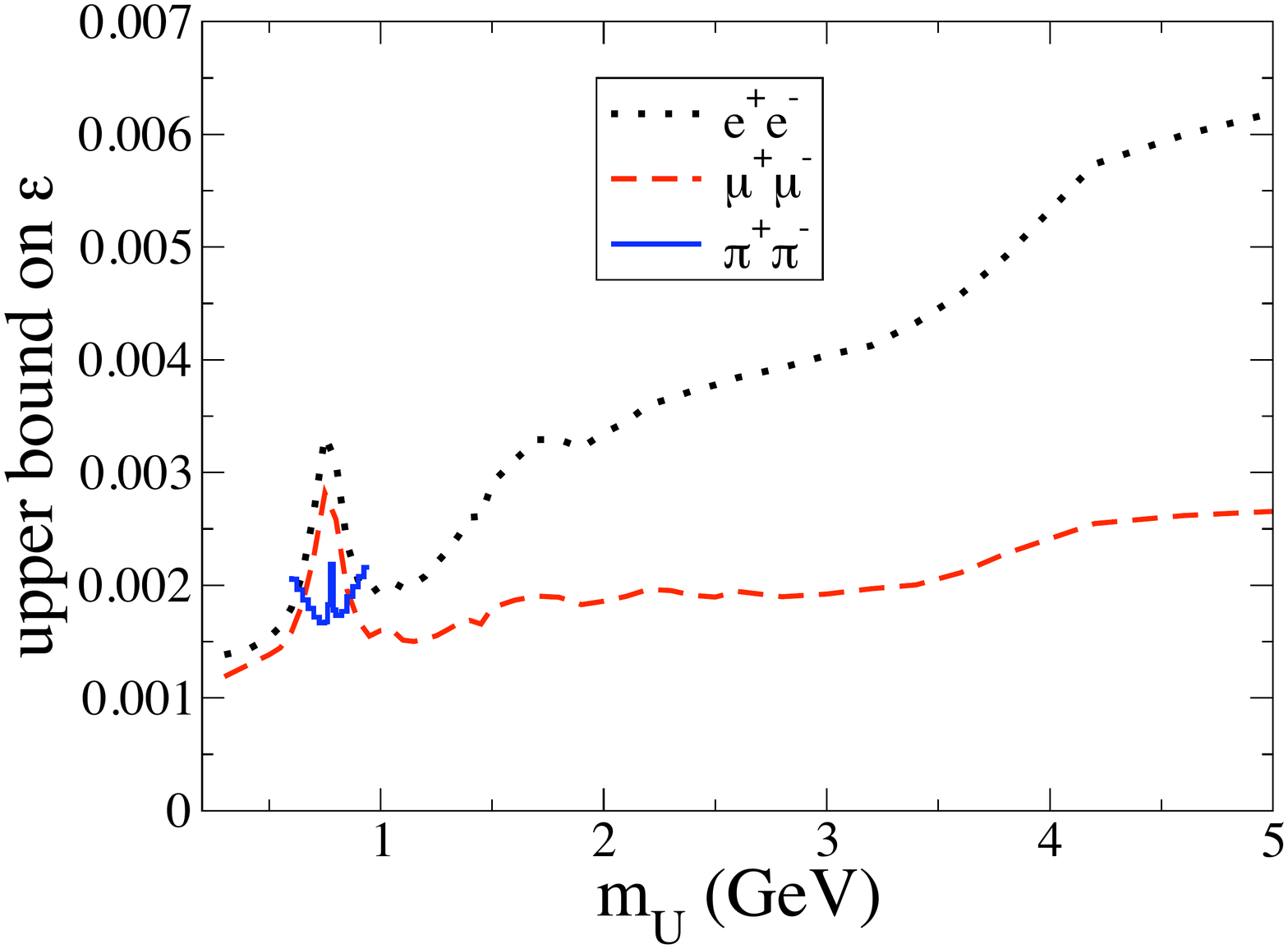}
\includegraphics[scale=0.27]{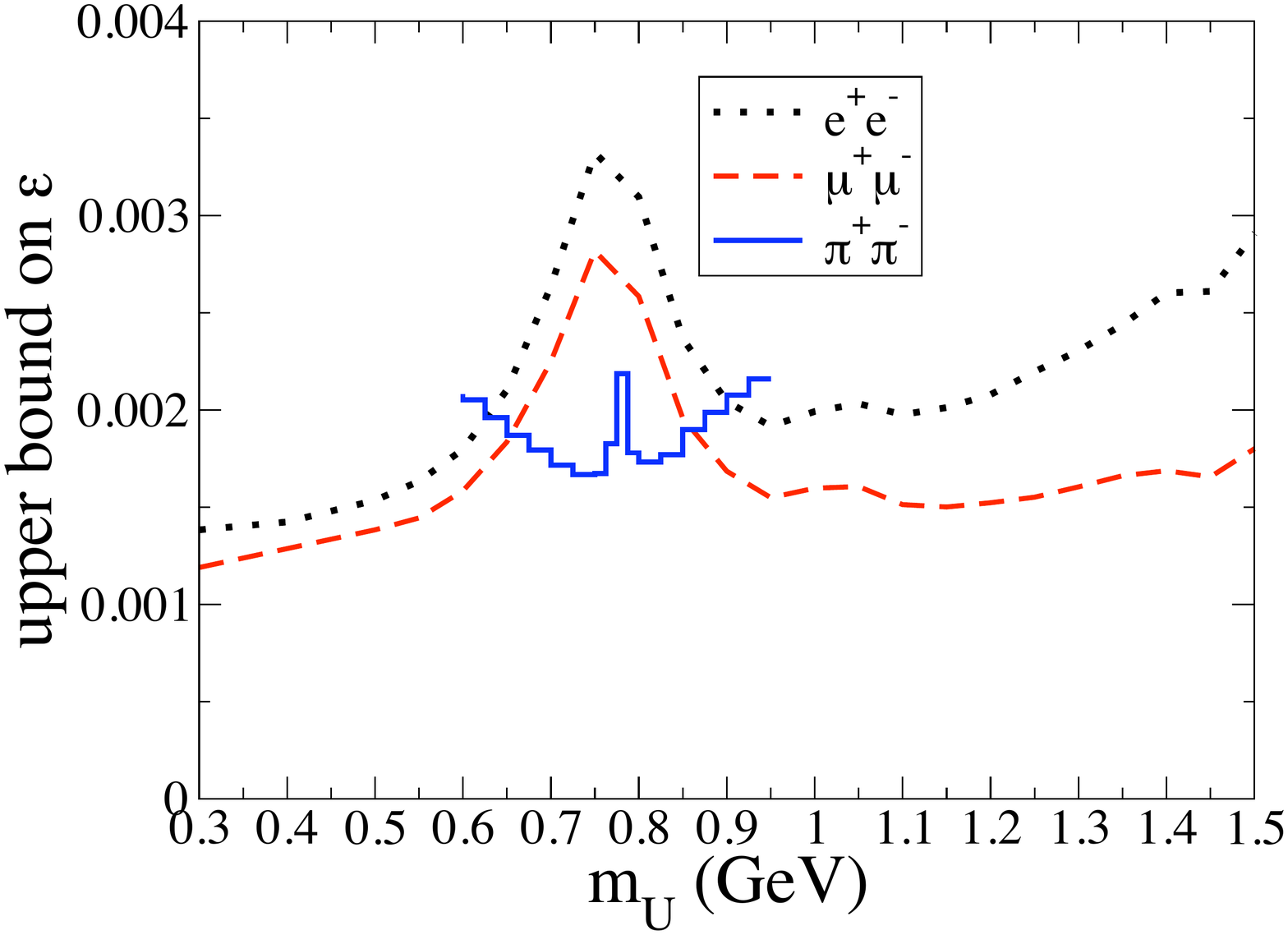}
\end{center} \caption{Reach at BaBar, defined to be the value of $\epsilon$ at which $S/\sqrt{B} = 5$, as a function of the $U$ mass. At left, the reach for $e^+e^-\gamma$, $\mu^+\mu^-\gamma$, $\pi^+\pi^-\gamma$,  channels; at right, the same figure with a zoom into the $m_U \leq 1.5$ GeV region. Note that muons are a much better search channel than electrons at larger $U$ masses, which is due to the large background from Bhabha scattering.} \label{fig:babarreach}
\end{figure}

The reach in $\ell^+\ell^-$ is significantly degraded if the U-boson mass is in the region of the $\rho$ resonance, because the branching fraction of $U \to \pi^+ \pi^-$ is enhanced in that region. We can recover some of the lost reach by doing a search in the $\pi^+ \pi^-$ channel for a peak. To compute the background, we use MadGraph with $\pi^\pm$ added as a new particle with modified interactions to take into account a form factor $F_\pi(q^2)$ for the pion-pion-photon vertex. Here $q^2$ is the virtuality of the photon, and we use the high-precision KLOE results for the form factor \cite{ee2pipi}. We will consider only the range $590~{\rm MeV} \leq m_U \leq 970~{\rm MeV}$. One could pursue similar studies at higher masses, but new hadronic decay modes open up and the analysis becomes more complicated. In the region we consider, nearly 100\% of the hadronic decays are to $\pi^+ \pi^-$ except on the $\omega$ resonance (mass 782 MeV, width 8.5 MeV). The $\omega$ decays to $\pi^+ \pi^- \pi^0$ 90\% of the time and to $\pi^0 \gamma$ 9\% of the time, so in this narrow region $\pi^+ \pi^-$ constitutes a relatively small fraction of all hadronic decays and the reach is significantly worse. (It could be improved by searching for the $\pi^+ \pi^- \pi^0$ final state only if the mass resolution achieved is significantly narrower than the $\omega$ width, which seems unlikely.)

In Figure \ref{fig:babarreach} we show the reach (i.e. that $\epsilon$ for which $S/\sqrt{B} = 5$).
Signal events are generated, for various choices of $m_U$, through a simple Monte Carlo based on the explicit formula for $d\sigma/d\cos\theta$ in $e^+ e^- \to U \gamma$ events followed by decay of the on-shell U-boson to $\ell^+ \ell^-$. The same acceptance cuts and smearing are applied. We then count the number of smeared signal events passing cuts, within a window of size $\delta m$, and compute $S/\sqrt{B}$.
It is important to know the branching ratios $Br(U \to e^+ e^-)$ and $Br(U \to \mu^+ \mu^-)$. For this, we have used the rate for $\gamma^* \to e^+ e^-$ versus $\gamma^* \to $ hadrons at $s_{\gamma^*} = m_U$, as extracted from the PDG tables of $R$ \cite{PDG-R}. We see from the figure that the $\mu^+ \mu^-$ channel has better reach for larger U-boson masses than the $e^+ e^-$ channel which suffers from a large Bhabha scattering background at higher $m_{e^+ e^-}$. 

We see from the figure that the reach in the $\mu^+ \mu^-$ channel slowly rises from $1.0 \times 10^{-3}$ at $m_U = 210$ MeV to about $2.7 \times 10^{-3}$ at $m_U=5$ GeV. The most noticeable exception is the window around $m_{\rho}$, where utilizing the $\pi^+ \pi^-$ mode helps to recover some the loss in the reach due to suppression of the leptonic branching ratio.
Note that although we take $S/\sqrt{B} = 5$ as our definition for the plot, a 5-sigma bump in a particular point in $m_U$ would {\em not} constitute a 5-sigma discovery, because the bump could occur at many different values of $m_U$. We won't explore the statistics of this ``trials factor" in any detail,
but caution the reader to take our 5-sigma estimate as an approximation of where first evidence could appear, not definitive discovery. 

For the integrated luminosity recorded at BaBar, we assume 432.89 fb$^{-1}$, the amount of data taken on the $\Upsilon(4S)$. We could include the full recorded BaBar data by simply scaling up the integrated luminosity to 531.43 fb$^{-1}$ \cite{BaBarLum}, but due to the one-fourth power scaling of reach with luminosity, this would amount to only a 5\% improvement.  Data collected at Belle, 895 fb$^{-1}$ \cite{BelleLum}, can also be combined. Since the kinematics and the resolution are similar, we can estimate its impact on the reach using luminosity. This would allow us to reach $\epsilon$ values about 25\% lower than those plotted.

Further small improvements might be obtained by also considering the QED process $e^+ e^- \to e^+ e^- U$. Although it is higher-order in QED, this process is not negligible because its cross section is strongly peaked at small values of the U-boson mass, due to the large forward enhancement from Bhabha scattering. The signal would be a narrow mass peak in two central leptons ($e^+ e^-$ or $\mu^+ \mu^-$), while the original $e^+$ and $e^-$ would be too forward to have reconstructable tracks in most events. Making the (optimistic) assumption of high trigger efficiency for these events, so that we only need to apply acceptance cuts on the lepton tracks, we estimate that the $e^+ e^- U(\to \mu^+\mu^-)$ channel offers a reach of $\epsilon \approx 2.7 \times 10^{-3}$ at $m_U \approx 250$ MeV, a little more than a factor of two worse than the $\gamma U(\to \mu^+ \mu^-)$ process considered above. At higher masses the reach compares even more unfavorably ($4.2 \times 10^{-3}$ at 500 MeV and $8.5 \times 10^{-3}$ at 800 MeV, for instance, more than a factor of three worse than $\gamma U$). 
In this estimate we have only taken into account QED contributions. $\Upsilon$ decays (for the 1S, 2S, and 3S) also contribute to this and similar channels, through $\Upsilon(1S) \to \ell^+ \ell^- U$. In the $\mu^+\mu^-\mu^+\mu^-$ channel, the $\Upsilon$ rate dominates: we expect that relative to QED it is enhanced by $R(e^+e^- \to \Upsilon) \times \BR(\Upsilon \to \mu^+ \mu^-) \approx (3 \times 10^3)\times(2\times 10^{-2}) = 60$. The enhancement of the signal rate in the $\Upsilon \to e^+ e^- U(\to \mu^+ \mu^-)$ channel relative to the QED process $e^+ e^- \to e^+ e^- U (U \to \mu^+ \mu^{-})$   is smaller due to the large forward enhancement of the latter. In any case, $60^{1/4}$ is not large. 
Moreover, such an estimate only applies if we are precisely on the resonance. In practice, due to the narrowness of $\Upsilon(1S)$,  $\Upsilon(2S)$,  and $\Upsilon(3S)$ resonances, there is always an additional suppression $\sim \Gamma_{\Upsilon} / \delta E_{\rm beam}\sim 10^{-2}$ \footnote{We would like to thank Maxim Pospelov and Adam Ritz for very useful discussion on this point.}. 
Thus, for the most part, we expect that four-lepton channels are less useful than $\ell^+\ell^- \gamma$. 
However, for small $m_U$, well below the di-muon mass threshold, the $e^+ e^- U(\to e^+ e^-)$ channel could be a useful complement to the $\gamma U(\to e^+ e^-)$ search, though such events could be difficult to trigger. Thus, we will leave more detailed consideration of this channel to the experiments.

Finally, we remark on a couple of ways in which our analysis can be improved. The degradation of the reach in the $e^+ e^-$ channel is due to the QED Bhabha scattering. We note that there should be clear angular correlation in the signal since $e^+ e^-$ are the decay products of a vector resonance. Since no such angular correlation is expected in the background, it provides one possible way to enhance the signal significance. We have used detector resolution as our choice of mass window. In practice, one should certainly optimize this choice to achieve maximal significance. 

\subsection{Similar Existing Searches, and the HyperCP Anomaly}

Both CLEO \cite{CLEOmumu} and BaBar \cite{BaBarmumu} have recently published searches for a light scalar $A^0$ decaying to $\mu^+ \mu^-$, arising in radiative decays $\Upsilon(1S) \to A^0 \gamma$ and $\Upsilon(3S) \to A^0 \gamma$ respectively. Thus the signal is a narrow peak in $\mu^+ \mu^-$ invariant mass in $\mu^+ \mu^- \gamma$ events, precisely of the same final states from continuum U-boson production. However, because they assume $A^0$ is a scalar arising in $\Upsilon$ decays, the analyses that have been performed do not translate directly to the limits we are interested in. In particular, in both cases data taken on the $\Upsilon(4S)$ (7 fb$^{-1}$ at CLEO, 78.5 fb$^{-1}$ at BaBar) were used to determine the background shape of $m(\mu^+ \mu^-)$ from continuum production. Data taken on a lighter resonance (1.1 fb$^{-1}$ or $(21.5 \pm 0.4) \times 10^6$ resonant decays on the $\Upsilon(1S)$ at CLEO; $122 \times 10^6$~$\Upsilon(3S)$ decays at BaBar) are then compared to the background shape, and limits are set on decays of the resonance to $A^0 \gamma$. The underlying assumption is that the $\Upsilon(4S)$ is so much broader than the narrower resonances that its branching fraction to $A^0\gamma$ would be negligible, and hence any $\mu^+\mu^-$ events appearing there are continuum. The final limits from BaBar are on the order of $10^{-6}$ for the product $Br(\Upsilon(3S) \to A^0 \gamma) \times Br(A^0 \to \mu^+ \mu^-)$. 

In these studies, the $A^0$ is assumed to be produced in decays, not by a QED continuum process, because it couples much more strongly to the $b$ quark than to electrons. For the vector U-boson we are interested in, continuum production is significant, because it couples to any electrically charged particle. Hence, we expect that if a peak were to show up in these searches, it would be most noticeable precisely in the larger $\Upsilon(4S)$ continuum data samples that have been used to determine the background shape! Nonetheless, we expect that these experiments are now well-equipped to do a careful version of the analysis we propose, using similar techniques to those used in these studies. It would require a bump-hunting analysis on the $\Upsilon(4S)$ data.

The BaBar analysis \cite{BaBarmumu} highlights a few experimental subtleties that we have glossed over above. One is the efficiency for positively identifying a muon (e.g., discriminating it from a pion). If only one muon is required to be cleanly identified, $\mu^+ \mu^-$ invariant mass on the $\Upsilon(3S)$ has a clearly visible $\rho$ peak, indicating that the sample is polluted by pions. If two IDed muons are required, there is a substantial loss of efficiency at low $m(\mu^+ \mu^-)$, but the $\rho$ peak disappears. We have assumed that all muons and pions are correctly identified. We don't expect that the efficiencies make a substantial difference in the reach, but this highlights that our rough estimates are only a guide to approximately how well we expect that the experiments can do. Similarly, the BaBar analysis finds two peaks with about 3$\sigma$ significance, which are not significant once the trials factor is taken into account. As we mentioned above, this indicates that our $5\sigma$ plots are only a guide to where first evidence might be expected.

In this context we should also mention the HyperCP result on the decay $\Sigma^+ \to p \mu^+ \mu^-$, which found three events clustered near $m(\mu^+\mu^-) = 214$ MeV \cite{HyperCPEvidence}. The event rate is compatible with Standard Model expectations, but the close clustering of masses led to suggestions that the decay could involve a light boson decaying to $\mu^+ \mu^-$. (In particular, the HyperCP collaboration estimated a 1\% probability for such clustering in the Standard Model, although the form factors for the decay have large uncertainties.) One goal of the CLEO and BaBar analyses was to constrain the existence of such a boson, if it is a pseudoscalar \cite{HyperCPSM,mangano_hypercp}. Recently, it has been argued that despite earlier claims that the only viable interpretations were a pseudoscalar or axial vector \cite{HyperCPSM}, the particular form of the U-boson's coupling (mimicking that of an off-shell photon) together with limited knowledge of form factors still allow for the possibility of a vector explanation of HyperCP \cite{PospelovSecluded}. However, this required $\epsilon^2 > 3 \times 10^{-5}$, somewhat larger than our expectations.

Although we can't directly interpret the CLEO or BaBar results as a limit on $\epsilon$, there appear to be no significant bumps in the plots available to us.
Our inability to strictly interpret such results as constraints on a U-boson should not prevent us from giving a rough estimate of how they limit $\epsilon$ if we interpret them favorably. Let's consider the CLEO study in slightly more detail, as it is encouraging that even this relatively small sample seems to be a good probe of the U-boson. Given that CLEO set a $2.3 \times 10^{-6}$ limit on the branching ratio $Br(\Upsilon(1S) \to A^0 \gamma)$ with $A^0 \to \mu^+ \mu^-$ 100\% of the time, in a sample of $21.5 \times 10^6$ $\Upsilon(1S)$ mesons, we can infer that there are no more than 49.5 signal events in the 1.1 fb$^{-1}$ sample. Interpreting the count as a direct limit on $\epsilon$, this translates to $\epsilon \lsim 7.5 \times 10^{-3}$. If we extrapolate to the full 8 fb$^{-1}$ sample, this would improve to $\epsilon \lsim 4.5 \times 10^{-3}$ (scaling by the one-fourth power of luminosity). We can arrive at a similar estimate by scaling our result on BaBar search in Sec.\ref{sec:babar-reach}. The estimated reach in continuum production at BaBar for an integrated luminosity of 433 fb$^{-1}$ is $\epsilon \approx 1.4 \times 10^{-3}$ for $m_U = 214.3$ MeV. Scaling this to the 8 fb$^{-1}$ at CLEO counting both $\Upsilon(1S)$ and $\Upsilon(4S)$, one obtains $\epsilon \approx 4 \times 10^{-3}$.  Hence we can't quite extrapolate from the lack of $\Upsilon(1S) \to A^0 \gamma$ signal at CLEO that the HyperCP U-boson is excluded, but we expect that the full $\Upsilon(1S) + \Upsilon(4S)$ sample does exclude HyperCP. A similar estimate suggests that the BaBar limit $Br(\Upsilon(3S) \to A^0 \gamma) \times Br(A^0 \to \mu^+ \mu^-) < 0.8 \times 10^{-6}$ (at the HyperCP mass) has already excluded the U-boson interpretation of HyperCP (and in fact sets a stronger bound of $\epsilon \lsim 2.0 \times 10^{-3}$). We hope that the collaborations will revisit their data and make this exclusion definitive. 

\section{$U$ in Meson Decays}
\label{sec:formulas}
\setcounter{footnote}{0}

\subsection{Generalities}

Low energy colliders also produced large numbers of mesons. Many mesons have decay channels into photons. Therefore, they can also decay into the U-boson with branching ratio $\BR (X\to Y+U) \approx \epsilon^2 \BR (X \rightarrow Y + \gamma)$. This is followed by $U \to \ell^+ \ell^-$. In cases where $\BR(X \to Y \ell^+ \ell^-) $ are measured and consistent with the SM prediction, a simple way to estimate the reach on the U-boson is to demand $\BR (X\to Y+U) \times \BR (U \to \ell^+ \ell^-)$ less than $x \BR (X \to Y \ell^+ \ell^-)_{\rm exp}$, where $x$ is determined by experimental precision and theoretical uncertainties.
However, the U-boson should be a narrow resonance in such channels. A dedicated search for it should have a better reach. In the following, we will first present estimates of discovery reach in various channels. Then, we perform a careful analysis for an important channel, $\phi \to \eta U$. The dominant background is $X \rightarrow Y + \gamma^* \rightarrow Y + \ell^+ \ell^- $, where $m_{\ell^+ \ell^-} = q_{\gamma^*} = m_U$. The number of events in a window of $\delta m$ around $m_{\ell^+ \ell^-} = m_U$ is approximately
\beq
n_{X} \BR (X\rightarrow Y + \ell^+ \ell^-) \frac{\delta m}{m_{U}} \frac{1}{\log [(m_{X} - m_Y)/2 m_{\ell}]},
\eeq 
where $n_X$ is the number of $X$ mesons which have been produced. (This assumes an approximate $\frac{1}{q_{\gamma^*}^2}$ dependence on the photon virtuality, which may be altered by form factors or by diagrams that contribute to the background but not the on-shell U-boson signal.)
The signal significance is 
\beq
\label{eq:meson-decay}
\frac{\mbox{S}}{\sqrt{\mbox{B}}} \approx \sqrt{n_X} \frac{\epsilon^2 \times \BR(X \rightarrow Y + \gamma)\times \BR(U \rightarrow \ell^+ \ell^-) }{\sqrt{\BR (X \rightarrow Y + \gamma^* \rightarrow Y + \ell^+ \ell^-)}} \sqrt{\frac{m_U}{\delta m} \log \left( \frac{m_X - m_Y}{2 m_{\ell}}\right)}.
\eeq
Typically, we have 
\beq
\BR (X \rightarrow Y + \gamma^* \rightarrow Y + \ell^+ \ell^-) \sim 10^{-2} \times BR(X \rightarrow Y + \gamma)
\eeq
Therefore, for percentage level branching ratio $\BR (X \rightarrow Y + \gamma)$, we need about $n_X \sim \mathcal{O}(10^{9})$ to reach $\epsilon \leq 10^{-3}$. Less statistics are required if the meson X has a larger branching ratio to photon.

\begin{table}
 \begin{tabular}{c|c|c|c|c|c}
\hline
\parbox{2cm}{$X \rightarrow Y U$} & $n_X$ &$m_X - m_{Y}$ (MeV) &$\BR (X \rightarrow Y + \gamma)$ &  $\BR (X \rightarrow Y + \ell^+ \ell^-)$ & $\epsilon \leq$ \\
\hline
$\eta \rightarrow \gamma U$ & $n_{\eta} \sim 10^7$&  547 & $2 \times 39.8 \% $ &  $6\times 10^{-4}$ & $ 2\times 10^{-3}$\\
$\omega \rightarrow \pi^0 U$ & $n_{\omega} \sim 10^{7}$& 648 & $8.9 \%$& $7.7 \times 10^{-4}$ &  $5 \times 10^{-3}$\\
$\phi \rightarrow \eta U$& $n_{\phi} \sim 10^{10}$ & 472& $1.3 \%$& $1.15 \times 10^{-4}$ & $1 \times 10^{-3}$ \\
$K_L^0 \rightarrow \gamma U$&  $n_{K^0_L} \sim 10^{11}$&  497 &  $2 \times (5.5 \times 10^{-4})$& $9.5 \times 10^{-6}$ & $2 \times 10^{-3}$\\
$K^+ \rightarrow \pi^+ U $& $n_{K^+} \sim 10^{10}$ & 354 & -  & $2.88 \times 10^{-7}$& $7 \times 10^{-3}$ \\
$K^+ \rightarrow \mu^+ \nu U $ & $n_{K^+} \sim 10^{10}$ & 392 &$6.2 \times 10^{-3}$ & $7 \times 10^{-8}$\footnote{Branching ratio $\BR (K^+ \to \mu^+ \nu e^+ e^-) $ for $m_{e^+ e^-} > $ 145 MeV \cite{kmunuee} } &  $2 \times 10^{-3}$\\
$K^+ \rightarrow e^+ \nu U $& $n_{K^+} \sim 10^{10}$& 496 &$1.5 \times 10^{-5}$ &$2.5 \times 10^{-8}$ & $7 \times 10^{-3}$\\
\hline
\end{tabular}
\caption{\label{tab:meson-decay} Reach in U-boson coupling in several competitive meson decay channels, assuming branching ratios to $e^+ e^-$, $\mu^+ \mu^{-}$ are similar if allowed by phase space. We take $m_U =250$ MeV for this table. $m_X - m_Y$ is the largest $m_U$ which can be probed in a particular channel, although reach will certainly reduce near kinematical boundary. Only $m_X - m_Y > 200 $ MeV included. We elaborate on the treatment of the Kaon decay channels and discuss the decays of $J/\psi$ and $\Upsilon$ in the text. Unless stated otherwise, the branching ratios are taken from the meson summary tables in Ref.~\cite{PDG}.}
\end{table}

In Table~\ref{tab:meson-decay}, we collect the set of potentially useful channels, and estimate their reach. The numbers $n_X$ are obtained either from meson factories in the cases of $\phi$ and Kaon, or estimated from the best measured decay branching ratios assuming $\sqrt{n}$ statistical fluctuations. As shown in Eq.~\ref{eq:meson-decay}, the coupling we probe scales as $\epsilon \propto n_X^{-1/4}$. Therefore, an order of magnitude uncertainty in estimating $n_X$ will result in about $25 \%$ error in the reach projection. 
We emphasize that the results presented here are only approximate order of magnitude projections, designed to highlight the useful channels. Accurate reach in each channel requires more careful analysis taking into account both theoretical considerations (such as form factors) and experimental details.

We briefly describe our treatment of the $K^+$ decay modes. The two body decay mode $K^+ \rightarrow \pi^+ \gamma $ with on-shell photon does not exist. We estimated the reach in this channel by assuming $\BR (K^+ \to \pi^+ U) \simeq \epsilon^2 \pi/\alpha \times \BR(K^+ \to \pi^+ e^+ e^-) $. The difference between $\BR (K^+ \to \mu^+ \nu \gamma)$ and $\BR (K^+ \to \mu^+ \nu e^+ e^-)$ cannot be fully accounted for by an additional factor of $\alpha/\pi$ and the fact that $K^+ \to \mu^+ \nu e^+ e^-$ has only been measured in the range of $m_{e^+ e^-}> 145 $ MeV \cite{kmunuee}. There is probably an additional factor of 10 from the form factor. To be conservative, we reduced signal rate by a factor of 10 in this channel. 

We discuss here other possible search channels, using the estimates shown in Table~\ref{tab:meson-decay} as rough guides for the size of branching ratios and total number of produced mesons for a certain channel to be competitive. In general, a search in a $V \to \gamma \ell^+ \ell^-$ channel would be quite similar to the continuum search discussed in Section \ref{sec:babar}. A challenge with such decays is discriminating them from the continuum QED process with initial-state radiation, so they have not been studied in much detail. We start with $J/\psi$. $J/\psi \to \gamma X$ is about a couple of per cent (not including $\to \gamma \eta_c(1S)$). We probably have $10^{7}$ $J/\psi$ \cite{besii_jpsi,cleo_jpsi,BaBarJPsi}. Therefore, by a rough comparison with $\phi$ decay,  this channel $J/\psi \to U + X$ is still probably at the level of $\epsilon \sim 10^{-2}$.  
Similarly, there is a channel $J/\psi \to U e^+ e^-$, to which the background process $J/\psi \to \gamma e^+ e^-$  has branching ratio $8.4 \times 10^{-3}$ \cite{JpsiEEGamma}, which again gives the same reach as the inclusive mode. The mode $J/\psi \to \gamma \mu^+ \mu^-$ has also been observed \cite{JpsiMuMuGamma}.   To be competitive with $\phi$, we probably need $>10^{10} J/\psi$, which will be achieved at BES-III \cite{besiii}. A search for a U-boson for in a lower mass range, $m_U \leq 50$ MeV, in the $U e^+ e^-$ channel has been considered in Ref.~\cite{zhu-bes}.

Next, we discuss $\psi(2S)$, of which about 14 million have been collected at the BES-II detector at BEPC \cite{moxh}. One possible interesting decay channel is $\psi(2S)\to \chi_{c0}\gamma$ at $\BR \sim 10 \%$. There is about 280 MeV of phase space. Therefore, it could be interesting to start to look for a low mass, $\sim 250$ MeV, U-boson in this channel. 

We also briefly discuss $\Upsilon$ decay. 
$\Upsilon(1S) \to \ell^+ \ell^- U$ could be a potentially interesting decay channel which can involve the U-boson, as we mentioned in the discussion of four-lepton final states in Sec. \ref{sec:babar-reach}. As far as we are aware the corresponding decays of the $\Upsilon(1S)$ have never been measured (though the $\phi \to \gamma \mu^+ \mu^-$ decay is observed with branching fraction at 10$^{-5}$ level \cite{PhiMuMuGamma}). $\Upsilon(4S)$ can be more interesting in this case. For example, $\Upsilon(4S) \to B B > 96 \%$ and $B \to D^0 / \bar{D}^0 + X \sim 62 \%$.   At the same time, the branching ratio $D^0 \to \eta + X \sim 10\%$. This can be an interesting source of $\eta$ mesons with $10^{8}-10^{9}$ $\Upsilon(4S)$. On the other hand, decay channels involving the photon, such as $B^0 \to \gamma + X \sim 10^{-7}$, are not useful at this stage. 

Finally, for completeness, we briefly remark on $\pi^0$ decay. From the best measured $\pi^0$ decay branching ratios, we can estimate probably close to $10^9$ $\pi^0$s have been studied in various channels. Based on our estimates, we see that it is in principle possible to probe $\epsilon$ down to $10^{-3}$ level for $m_{\rm U} < m_{\pi}$ in $\pi \to \gamma U$. Since the signal is $\gamma e^+ e^-$, such detection may have lower sensitivity due to larger backgrounds (see also \cite{gninenko}). The U-boson induced correction to the decay branching ratio $\BR (\pi \to e^+ e^-)$ has also been considered \cite{dorokhov}. However, this correction is proportional to the axial couplings of U-boson to SM fermions. The U-boson considered here only acquires axial coupling through mixing with the Z, which suffers from an additional suppression of $m_U^2/m_{Z}^2$.  Therefore, this channel will allow $\epsilon \sim 1$. 

We have discussed signals in which a photon can be replaced by a U-boson. One can also consider signals involving the dark Higgs, if it is kinematically accessible. Such signals would be precisely the Higgs$'$-strahlung process discussed in Ref. \cite{Batell:2009yf}, but through a production mechanism like $e^+ e^- \to J/\psi~({\rm or}~\Upsilon) \to U h_d$ rather than simply through an off-shell U-boson. However, in comparison with the process studied in Ref.~\cite{Batell:2009yf},  resonant production will suffer from an additional suppression of order $10^{-2}$ due to the spread in beam energy, as we have discussed in Section \ref{sec:babar-reach}. 

\subsection{Searching in $\phi$ Decays}
\label{sec:kloe}
\setcounter{footnote}{0}

The KLOE experiment has collected about 2.5 fb$^{-1}$ of luminosity, which amounts to about 8 billion $\phi$ mesons \cite{KLOErecent}. The $\phi$, with a mass of 1019.3 MeV, decays to $\eta \gamma$ 1.3\% of the time; the $\eta$ (mass 547.9 MeV) decays to $\gamma \gamma$ 39\% of the time. Thus we can search for $\phi \rightarrow \eta U$ at KLOE, or $\phi \rightarrow \eta \gamma$ followed by $\eta \rightarrow \gamma U$. Momenta of charged tracks at KLOE are measured to about $0.4\%$ accuracy \cite{KLOEreview}.

The decays we will be interested in will be those which arise from Standard Model $P \to V \gamma$ or $V \to P \gamma$ processes ($P$ meaning ``pseudoscalar" and $V$ ``vector"), where we replace the $\gamma$ by the new gauge boson $U$. The irreducible SM background will be the process where the $\gamma$ is offshell: $\gamma^* \to \ell^+ \ell^-$. The effective interaction is:
\beq
{\cal L}_{VP\gamma} = f_{VP\gamma} \epsilon^{\mu \nu \rho \sigma} \partial_\mu A_\nu \partial_\rho V_\sigma P.
\eeq
We will multiply by a form factor $F_{VP\gamma^*}(q^2)$ to model the interaction with an off-shell photon (and, through this, with the new gauge boson $U$), where $q^2$ is the photon virtuality and the normalization is $F_{VP\gamma^*}(0) = 1$. Detailed discussion of form factors for many processes of this type can be found in Ref. \cite{LandsbergReview}.

In general $V \to P$ transitions the differential rate to $\ell^+ \ell^-$ versus the rate to on-shell $\gamma$ is:
\beq \label{eq:Pll}
\frac{d}{dq^2} \frac{\Gamma(V \to P \ell^+ \ell^-)}{\Gamma(V \to P \gamma)} & = & \frac{\alpha}{3\pi} \frac{\left|F_{VP\gamma^*}(q^2)\right|^2}{q^2} \sqrt{1-\frac{4 m_\ell^2}{q^2}} \left(1 + \frac{2 m_\ell^2}{q^2}\right) \frac{\lambda^{3/2}(m_V^2, m_P^2, q^2)}{\lambda^{3/2}(m_V^2, m_P^2, 0)},
\eeq
so that if we are looking near invariant mass $q^2$ with resolution $dq^2$ the relative rate is $\frac{\alpha}{3\pi} \frac{dq^2}{q^2}$, up to phase-space factors and an order-one form factor. On the other hand, the rate for production of an on-shell, narrow $U$ gauge boson is:
\beq
\label{eq:UvsA}
\frac{\Gamma(V \to P U)}{\Gamma(V \to P \gamma)} = \epsilon^2 \left|F_{VP\gamma^*}(m_U^2)\right|^2 \frac{\lambda^{3/2}(m_V^2, m_P^2, m_U^2)}{\lambda^{3/2}(m_V^2, m_P^2, 0)}.
\eeq
(Here $\lambda$ is the usual kinematic function $\lambda(x,y,z) = (x-(\sqrt{y}+\sqrt{z})^2)(x-(\sqrt{y}-\sqrt{z})^2)$.) 

\begin{figure}[!h]
\begin{center}
\includegraphics[width=7cm]{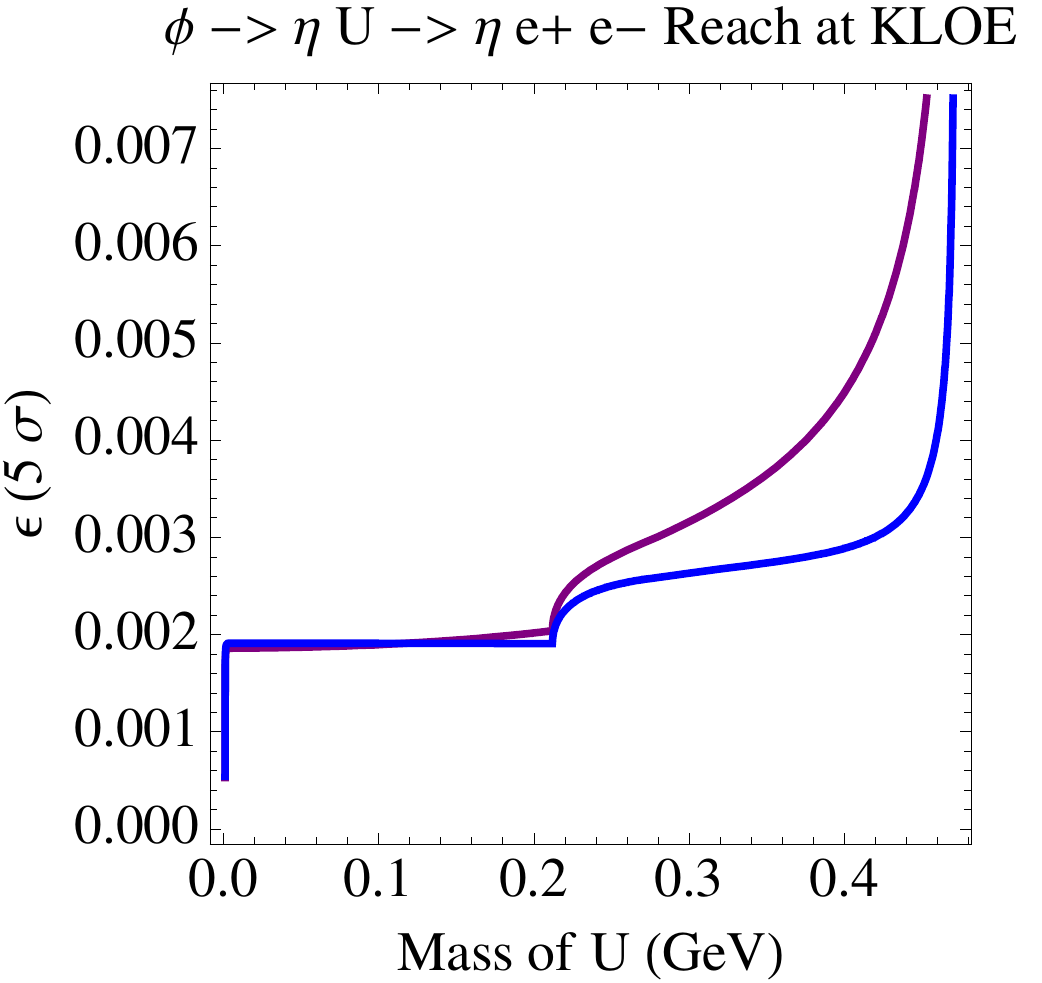}\includegraphics[width=7cm]{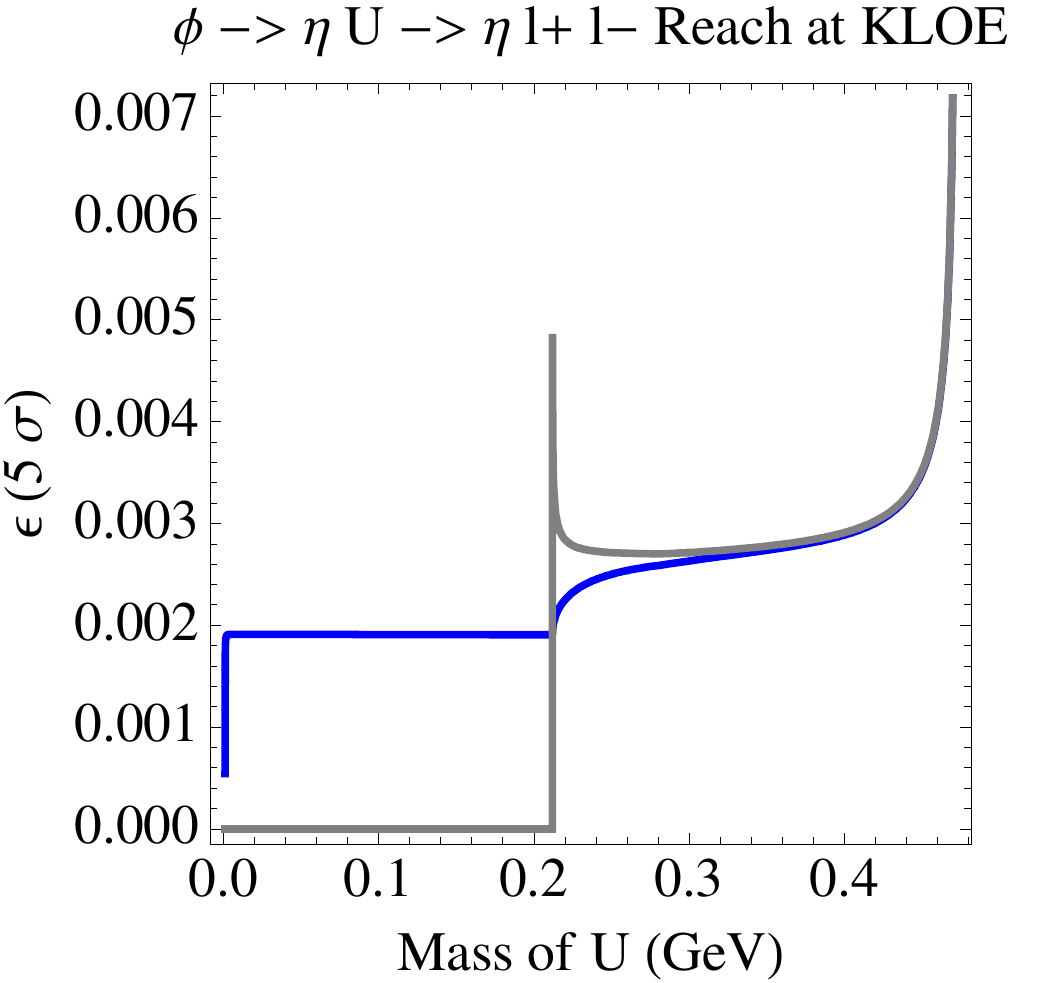}
\end{center} \caption{Reach for $U \to e^+ e^-$ at KLOE in the process $\phi \to \eta U$.  {\bf At left}: reach in $e^+e^-$. The upper (purple) curve is for constant form factor $F_{\phi\eta\gamma^*}(q^2) = 1$, whereas the lower (blue) curve is for the single-pole fit $F_{\phi\eta\gamma^*}(q^2) = 1/(1 - 3.8 {\rm GeV}^{-2} q^2)$ from Ref. \cite{Achasov:2000ne}. {\bf At right}: The blue curve is the reach in $e^+ e^-$ (with single-pole form factor fit) and the gray curve is the corresponding reach in $\mu^+ \mu^-$.} \label{fig:KLOEreach}
\end{figure}

In Figure \ref{fig:KLOEreach} we present the reach of the U-boson using $\phi \to \eta U$. Specifically, we estimate the number of background events at given $m_U^2$ by taking the shape of the background distribution from Eq. \ref{eq:Pll} and normalizing it to the PDG branching fraction times 8 billion total $\phi$'s. Given this number of background events (in a window of size given by KLOE's resolution on the mass), Eq. \ref{eq:UvsA} was used to estimate the corresponding number of signal events, and the value of $\epsilon$ for which $S/\sqrt{B} = 5$ is plotted. The KLOE momentum resolution of 0.4\% is used (here we neglect smearing in the angular directions, which we do not expect to significantly alter the results). The 5$\sigma$ reach at 214 MeV is slightly worse than our back-of-the-envelope estimate: about $2 \times 10^{-3}$. Different choices of form factor significantly modify the reach near the upper end of the kinematically accessible range for $m_U$, but have little effect at the low end. The kink in the curve just above 0.2 GeV is from the sudden drop in $U \to e^+ e^-$ branching fraction when the $\mu^+ \mu^-$ decay mode opens up. At around 0.28 GeV, the decay to $\pi^+ \pi^-$ opens up. We estimate its branching ratio using a vector meson dominance model of the pion form factor, $F_\pi(q^2) = \frac{1}{1 - q^2/m_\rho^2}$, which is approximately valid since we are probing $m_U \ll m_\rho$. The branching fraction to pions remains small ($\approx 14\%$) even at $m_U \approx m_\phi - m_\eta$, so this mode has only a small effect on the reach.

After the di-muon threshold, the combined reach of $e^+ e^-$ and $\mu^+ \mu^-$ is similar to the $e^+ e^-$ reach below the threshold.

\section{Fixed Target Experiments}
\label{sec:fixed}
\setcounter{footnote}{0}

Let's consider a different option: in a fixed target experiment, instead of trying to produce the U-boson in scattering or in a rare decay, we can let electrons propagate through some length of material. Rarely, an electron-proton interaction will produce a U-boson. We won't be able to isolate electrons in an EM shower well enough to resolve a peak in $m(e^+e^-)$, but if we have good muon identification and momentum resolution we might hope to place a muon detector outside the target and find the $U$ mass peak in $\mu^+ \mu^-$. In this section we will examine whether any existing fixed target experiments have the capability to discover the U-boson, and estimate the properties we would want to have in an ideal fixed target $U$ search experiment.

We consider the process $e^- p \to e^- U p$, followed by $U \to \mu^+ \mu^-$.
Electrons are incident on a fixed target at a rate of $\dot{n}_e$. For an electron beam with $n_e$ electrons per bunch and bunch spacing $\tau$, $\dot{n}_e=n_e / \tau$. The target, with thickness $L$ and density $\rho$,  is made of material of atomic number $Z$ and atomic weight $A$. The instantaneous luminosity of electron-proton collisions is
\beq
\frac{d{\cal L}}{dt} & = & \dot{n}_e  n_{X_0} X_0  N_A (Z/A),
\eeq
where $X_0$ is the radiation length. The number of radiation lengths of the target $n_{X_0}$ is determined by the thickness of the target as $\rho L = n_{X_0} X_0$. $N_A = 6.02\times 10^{23} {\rm atom}/{\rm mol}$ is Avogadro's number. Radiation lengths $X_0$ in some materials are: helium, 94.3 $g$ cm$^{-2}$;  iron, 13.8 $g$ cm$^{-2}$; lead, 6.4 $g$ cm$^{-2}$. For an idealized example, consider an electron beam similar to that used at BaBar, with $2.1 \times 10^{11}$ electrons per bunch and bunch spacing $\tau=4.2$ ns.\footnote{One reason this is idealized is that bunches at BaBar can circle the ring repeatedly, whereas a bunch incident on a fixed target is not so easily recycled. Hence delivering a similar number of electrons at such short bunch spacing on a fixed target is probably unrealistic. As we will discuss, readout rates are another limiting factor, and by the end of this section we will arrive at a more realistic estimate.} We  can achieve an instantaneous luminosity of $n_{X_0} \times 1.9 \times 10^{43}$ cm$^{-2}$ s$^{-1}$ on an iron target. This is certainly a large enhancement in comparison with the instantaneous luminosity on the order of $10^{33}$ to $10^{34}$ cm$^{-2}$ s$^{-1}$ at BaBar. 

\begin{figure}[!h]
\begin{center}
\includegraphics[width=12cm]{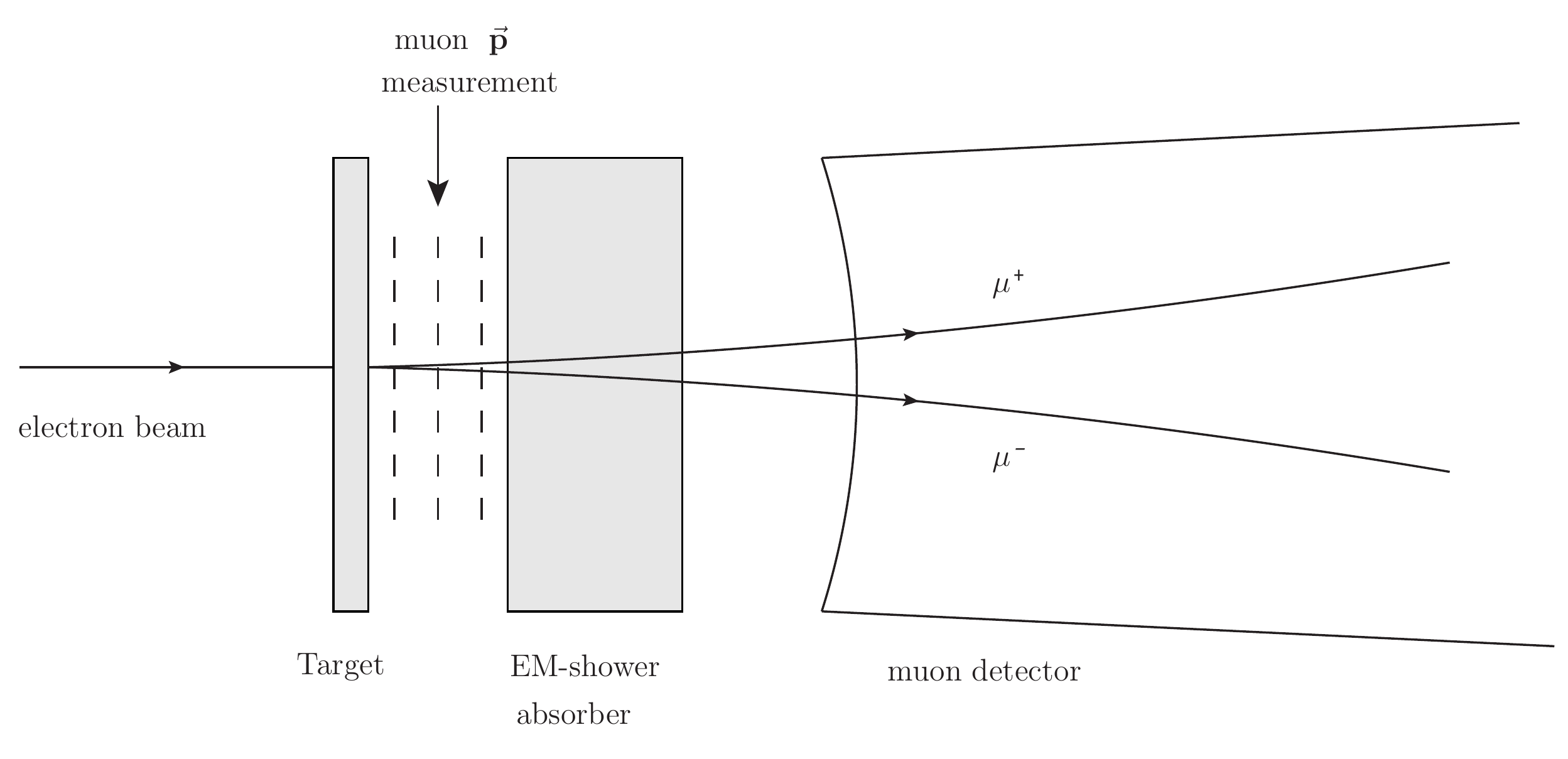}
\end{center} \caption{Schematic of a possible fixed-target experiment to search for the U-boson. An electron beam impacts a fixed target, with beam intensity and target thickness designed to produce about one $\mu^+ \mu^-$ pair per bunch. Tracks emerging from the target are measured in a magnetic field. An absorber stops electrons and photons emerging from the target, while a more distant detector tags the muons.} \label{fig:fixtarget}
\end{figure}

Treating the proton approximately as a point particle \footnote{The process under consideration produces U-bosons moving forward with approximately the beam energy. When the beam energy is much larger than $m_U$, the momentum exchanged in the process is suppressed below $m_U$. The momentum exchange at which scattering begins to resolve nuclear substructure is set by the nuclear radius $R \approx A^{1/3}~{\rm fm}$ (e.g., $R^{-1}$ is about 50 MeV for iron). Thus the regime where the electron scatters coherently off the whole nucleus, and $\sigma \propto Z^2$ rather than $\propto Z$, is more important than that where the substructure of an individual proton is resolved. For the rough order-of-magnitude estimates we do here, we will neglect this factor. (The change in reach is $Z^{1/4}$ at fixed luminosity, which is not a large factor.) Note also that for larger masses, the cross section scales as $1/m_{\rm U}^2$, in addition to form-factor effects. After the initial version of this paper, a detailed calculation of production processes and a discussion of various fixed-target designs has appeared in Ref. \cite{Bjorken:2009mm}. We refer the reader there for detailed discussion of nuclear form factors, the Weisz\"acker-Williams approximation to production, and many related issues.}, the signal rate can be estimated to be $\sigma_s^{\rm ep} = \epsilon^2 \sigma_0^{\rm ep} \sim$~1~pb~$=$~10$^{-36}$ cm$^2$ for $\epsilon \sim 3 \times 10^{-3}$. $\sigma_0^{\rm ep} = 10^{-30}$ cm$^{2}$ is the cross section for photon production $e^- p \to \gamma e^- p$.  The background is $e^- p \to e^- \mu^+ \mu^{-} p $ through an off-shell photon, with a rate
$\frac{\alpha}{\pi} \sigma_0^{\rm ep} \times \frac{d \cal L}{dt},$
which is much bigger than the signal.
Note that for this process in a fixed-target experiment, unlike the processes we considered at BaBar, there is a strong enhancement in the rate because we can allow the $U$ boson to be forward.
 In particular, on an iron target, there will be on the order of 
\beq
\left(\frac{n_{\mu^+ \mu^- }^{\rm bkgd}}{\text{bunch}}\right) \sim 10 \times n_{X_0} \times \left(\frac{n_e}{2 \times 10^{11}} \right)
\left(\frac{X_0}{13.8 \text{ g}\text{ cm}^{-2}} \right)
\left( \frac{Z}{26} \right) \left( \frac{55.8}{A}\right).
\eeq
background $\mu^+ \mu^-$ pairs per bunch. The high combinatorial background, $\sim (n_{\mu^+ \mu^- }^{\rm bkgd})^2$, can seriously degrade the signal. We can choose to either use a less intense beam $n_e \sim 10^{10}$, or a thinner target $ n_{X_0} \sim 0.1 $ (about $0.18$ cm for an iron target) to effectively achieve $n_{\mu^+ \mu^- }^{\rm bkgd} < 1$. 
In the case of a thinner target,  one would want something resembling a typical collider detector outside the material: we must be able to measure tracks from charged particles emerging from the material. Then a calorimeter will stop the many electrons and photons that emerge, while muons pass through to some further detector which can ID them as muons and allow matching to the track measured in the inner detector. This matching is complicated by the fact that muons passing through the absorber will change direction due to their interactions with the material, and there are many candidate tracks that they might match to. This is potentially a limiting factor on the resolution\footnote{We thank Tom LeCompte and Henry Lubatti for this observation.}  (and may prove to be the largest experimental difficulty in this setup), but we will leave a more thorough consideration to future work.We show a schematic of the experimental setup in Figure~\ref{fig:fixtarget}.

Muons passing through dense material will lose energy due to ionization, which can degrade the resolution. The Bethe-Bloch estimate for energy lost by a minimum-ionizing muon from ionization in a medium is, roughly, 
\beq
\delta E \sim X_0 \times \left(\frac{dE}{dx}\right)_{\rm min} \approx 1.5~ {\rm MeV}~{\rm cm}^2~{\rm g}^{-1} \times X_0,
\eeq
and muons at the energies we consider are minimum-ionizing to good approximation. There is only weak
dependence on the material. For instance, in helium gas,
$\left<-dE/dx\right>_{\rm min} = 1.937$ MeV cm$^2$ g$^{-1}$; in liquid
argon, 1.502 MeV cm$^2$ g$^{-1}$; in iron, 1.451 MeV cm$^2$ g$^{-1}$;
in lead, 1.122 MeV cm$^2$ g$^{-1}$ \cite{MuonStopping}. As we have seen, energy resolution of a typical muon detector is $\delta E \sim $ several MeV. If we require the degradation of resolution due to ionization not to be worse than the detector resolution, we would use target of thickness less than $n_{X_0} \sim 0.1$, which is in the same range as we obtained by requiring no background pile-up. 

With no pile-up (i.e. no more than one expected $\mu^+ \mu^-$ pair per bunch), the signal significance of a fixed target experiment is
\beq
\frac{\text{S}}{\sqrt{\text{B}}} \simeq \epsilon^2 \times \BR (U \to \mu^+ \mu^-)  \times \sqrt{\frac{d \cal L}{d t} \Delta T}  \times \sqrt{\sigma_0^{\rm ep}} \times \sqrt{\frac{\pi}{\alpha} \frac{m_U}{\delta m_{\mu^+ \mu^-}}},
\eeq
where $\Delta T$ is the run time. 
$\delta m_{\mu^+ \mu^-}$ is proportional to the thickness of the material if the ionization energy loss dominates the error in muon momentum measurement. In this case, the signal significance is not sensitive to the thickness. However, as we have argued above, requiring no pile up already restricts us away from this regime, and the resolution will be determined by the muon detector. In this case, the signal significance is proportional to the square root of the thickness. Supposing we use an iron target with thickness $n_X = 0.1$ (0.18 cm), a  BaBar-like electron beam, and $\Delta T$ similar to BaBar run time, we can expect an improvement up to $\epsilon \sim 10^{-6}$.  

However, another important limitation can come from  the maximal event recording rate, $R_{\rm max}$. For detectors at modern high energy colliders, typically, $R_{\rm max} \sim 10^2$ Hz. However, the read-out rate at lower energy detectors can be as high as $10^5$ Hz (see e.g. Ref. \cite{Alcorn:2004sb}). In principle, since we require much less information, basically only invariant mass $m_{\mu^+ \mu^-}$, to carry out the dedicated search for the U-boson, further significant improvements are conceivable. We would hope that a significant portion of the analysis could be carried out in hardware for maximum rate.

The setup can only be designed so that the rate of QED background $e^- p \to e^- p \gamma^*~\to~e^- p \mu^+ \mu^-$, is not larger than $R_{\rm max}$, since any pre-scaling will lead to worse degradation of the signal. In this case, this signal rate is
\beq
R_{\rm signal} \simeq \frac{\pi \epsilon^2}{\alpha}R_{\rm max},
\eeq
and the significance is
\beq
\frac{\text{S}}{\sqrt{\text{B}}} &\simeq& \frac{R_{\rm signal}}{\sqrt{R_{\rm max}}} \sqrt{\frac{m_{U}}{\delta m}} \sqrt{\Delta T} \nonumber \\
&=& \frac{\pi \epsilon^2}{\alpha} \sqrt{\frac{m_{U}}{\delta m}} \sqrt{R_{\rm max}} \sqrt{\Delta T}. 
\eeq
Therefore, the required running time $\Delta T$ for some desired precision is
\beq
\frac{\Delta T}{\text{year}} \simeq \frac{1}{\epsilon^4} \frac{1}{10^{12}} \left( \frac{\text{Hz}}{R_{\rm max}} \right) \left( \frac{\delta m}{m_U}\right).
\eeq
We see that for $R_{\rm max} \sim 10^5$ Hz, which corresponds to an instantaneous luminosity on the order of 10$^{38}$ cm$^{-2}$ s$^{-1}$, we can reach to about $\epsilon \sim 10^{-5}$. We have dropped a number of order-one factors, not to mention experimental efficiencies, so conservatively one should interpret this as a few $\times 10^{-5}$.

As a sanity check on these numbers, we should consider the luminosity that has been achieved at existing fixed-target experiments. For instance, a SLAC beam-dump experiment that searched for millicharged particles involved $3 \times 10^{10}$ electrons on 6 radiation lengths of a tungsten/rhenium target at a rate of 120 Hz \cite{Prinz:1998ua}. This corresponds to an instantaneous luminosity on the order of $10^{37}$ cm$^{-2}$ s$^{-1}$. In fact, this experiment already can set interesting limits on certain U-boson scenarios \cite{BeamDumpUBoson}. A recent study of deeply virtual Compton scattering at JLab Hall A used a 5.75 GeV electron beam on a liquid $D_2$ target, with $4 \times 10^{37}$ cm$^{-2}$ s$^{-1}$/nucleon luminosity \cite{Mamouz:2007vj} . Experiments involving parity-violating electron scattering from nuclei at MIT-Bates, JLab, and Mainz have used luminosities greater than 10$^{38}$ cm$^{-2}$ s$^{-1}$, as reviewed in Ref. \cite{Beck:2006ac}. In particular, JLab Hall A can reach a luminosity of $5 \times 10^{38}$ cm$^{-2}$ s$^{-1}$ \cite{Alcorn:2004sb}. Based on these examples, we conclude that a luminosity 10$^5$ times higher than that of BaBar is not unrealistic.

We also briefly comment on other possible beams. Because the only prerequisite for producing a U-boson is an electromagnetically interacting particle, we expect that proton or pion beams incident on fixed targets can also produce U-bosons, with cross sections comparable to those obtained with electron beams. However, the larger total $\pi p$ and $p p$ interaction cross-sections could lead to more difficulty isolating a clean signal. Furthermore, fixed target experiments with hadronic beams typically have lower luminosities than we require. For instance, the charmonium experiment E835 at FNAL used an antiproton beam and recorded typical instantaneous luminosities of $2 \times 10^{31}$ cm$^{-2}$ s$^{-1}$ \cite{E835}. As another example, the proton beam used at HyperCP involved only on the order of 10$^{10}$ protons per second \cite{HyperCP}. Hence, we expect that the very high luminosities we require are most easily achieved with electron beams.

\subsection{Long Lifetimes}

\begin{figure}[!h]
\begin{center}
\includegraphics[width=8cm]{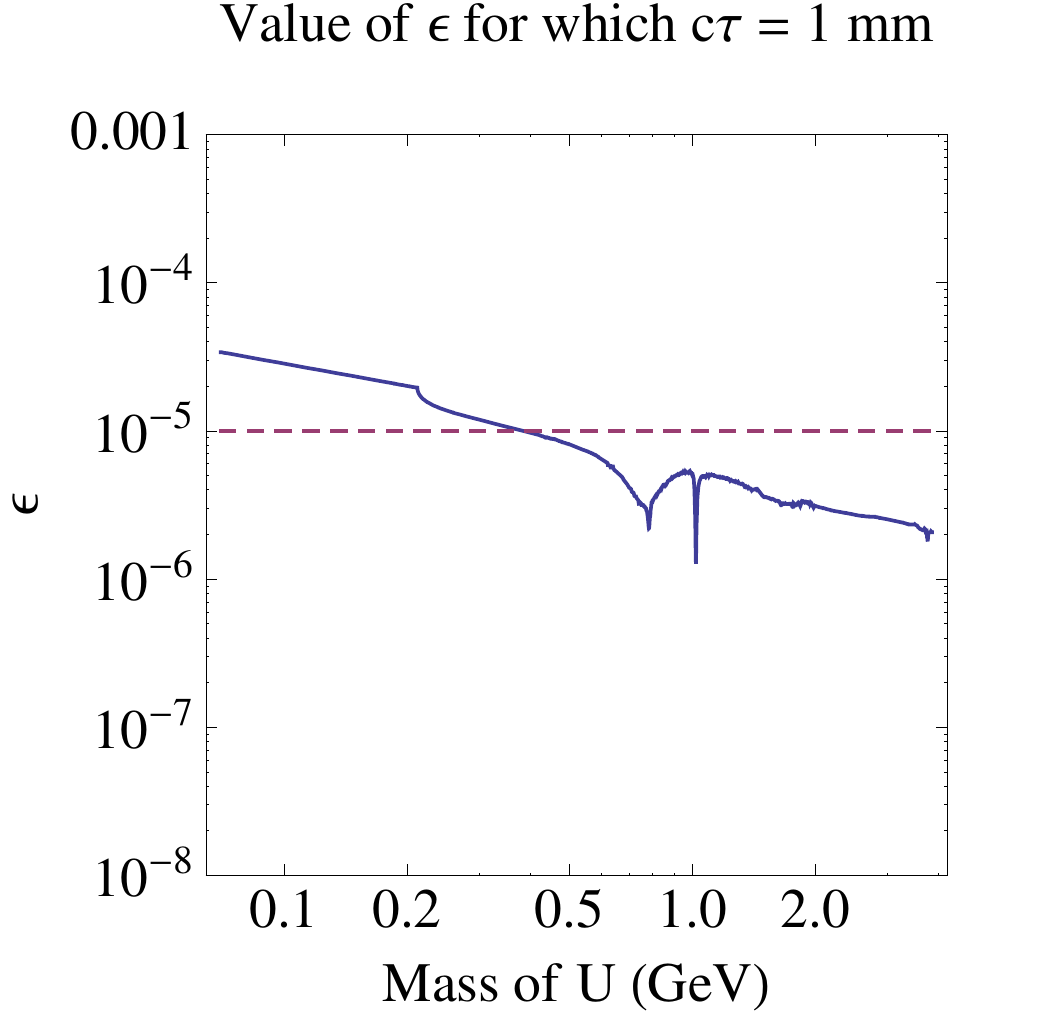}
\end{center} \caption{The value of $\epsilon$ at which the decay length $c\tau$ becomes 1 millimeter, as a function of the U-boson mass $m_U$. The dashed horizontal line at 10$^{-5}$ indicates roughly the lowest $\epsilon$ we could hope to probe with a fixed-target experiment designed for prompt decays.} \label{fig:lifetime}
\end{figure}

In the above discussion, we have argued that a dedicated fixed target experiment can reach values of $\epsilon$ on the order of $10^{-5}$. At such small values of $\epsilon$, it becomes interesting to consider the lifetime of the U-boson, which may travel a macroscopic distance before decaying. (Parametrically, the decay length $c\tau \sim \epsilon^{-2} \frac{\pi}{\alpha m_U}$.) We gain new reach by observing that Standard Model background will consist of muons originating within the target while the decay of the U-boson can happen within the tracking system. We have discussed using a target with thickness on the order of 0.1 radiation lengths, which for iron is 1.8 mm. Thus lifetimes $\gamma c\tau$ on the order of millimeters to centimeters or longer could potentially be probed by reconstructing tracks that originate outside the target. The boost of the U-boson, $\gamma \sim E_{\rm beam}/ m_{U} \gg 1$, improves the prospects (so in particular, more energetic beams can help produce more highly displaced vertices). In Figure~\ref{fig:lifetime}, we plot the value of $\epsilon$ at which $c\tau = 1~{\rm mm}$ as a function of the U-boson mass. At low masses this is $\epsilon \sim 10^{-5}$,  falling closer to $10^{-6}$ as the U-boson mass approaches 1 GeV. Below this value of $\epsilon$, we expect that the Standard Model background becomes very low. One source is conversion of real photons to $\mu^+\mu^-$ pairs in material, so one would want a tracker containing very few radiation lengths of material (the spatial distribution of vertices could perhaps also help distinguish conversions from decays). There would also be some chance of tracks being misreconstructed to produce fake displaced vertices. These effects can only be estimated given a specific experimental design. Depending on the experimental details, $\epsilon$ below 10$^{-7}$ could be reached in this way. In this way experiments could potentially cover the entire interesting range of $\epsilon$ for $2 m_\mu \lsim m_U \lsim 1$ GeV, but as $m_U$ grows closer to 1 GeV it is possible that there is a window of $10^{-6} \lsim \epsilon \lsim 10^{-5}$ that might not be reached by the methods we propose.

\section{Other possible probes of the U-boson}

\subsection{High-Energy Colliders}
\label{sec:highenergy}
\setcounter{footnote}{0}

Here we will briefly consider the production of the $U$ boson at high-energy (i.e. order 100 to 1000 GeV) collider experiments. First, note that any process we consider will involve the same comparison of a signal rate $S \approx \epsilon^2 N$, where $N$ is roughly the number of events in a corresponding Standard Model process involving photons, with a background rate near a given bin $B \approx \frac{\alpha}{\pi} \frac{\delta m}{m_U} N$. Such an estimate tells us that to probe $\epsilon \approx 10^{-3}$, we always want $N$ to be of order one billion or more.

The first high-energy collider to consider is LEP. At LEP, the event rate was simply too small. LEP 1 recorded a total of 17 million $Z$ bosons among all of the experiments (corresponding to approximately 200 pb$^{-1}$/experiment), whereas LEP 2 delivered 200 pb$^{-1}$ of luminosity at each of 7 different energies from 189 GeV to 207 GeV, where the cross section is much smaller than on the $Z$ pole \cite{LEPrefs}. From this we see that combining all of LEP's data, even before focusing on events containing a photon, we do not achieve $N \gsim 10^9$.

There is a Z rare decay channel, $\BR \sim \epsilon^2$, into the dark sector. Depending on the model, it can either result in 1 lepton jet $+ \not{\!\!E}_{T}$, or a pair of lepton jets and possibly additional $\not{\!\!E}_{T}$ \cite{models}. Given that LEP has produced about 17 million Zs, there should be a possible signal or constraint for the coupling roughly on the order of $\epsilon \sim 10^{-3}$ already, if this particular configuration is searched for. 

Next we consider the Tevatron, where the two experiments have recorded about 10 fb$^{-1}$ of total data. With an average instantaneous luminosity of around $2 \times 10^{32}$ cm$^{-2}$s$^{-1}$ and a trigger rate of about 50 Hz \cite{CDFtrigger, D0trigger}, we estimate that the two experiments together have around $5 \times 10^9$ total events on tape, and will add billions more by the end of their runs. Most of this rate does not involve events with isolated photons. The prompt central photon rate at the Tevatron is about 30 pb/GeV per unit rapidity at $p_t^\gamma \approx 30$ GeV \cite{D0photons}, increasing toward smaller $p_t$. We can crudely estimate that the rate for photons with $p_t$ between 20 and 35 GeV is on the order of several nb, with perhaps on the order of 10$^7$ events on tape. Hence the analogous process with a $U$ boson produced in this range is not very useful for probing $\epsilon$ on the order of $10^{-3}$. Very soft photons are more plentiful, but one typically needs some harder object in the event to trigger on.

On the other hand, there is still some hope of an interesting search at the Tevatron. We have on the order of a billion events to work with, and most of the events that have triggered will contain some jet activity. Inside these jets, there are numerous charged particles. Just as a high-energy quark can radiate not only gluons but also photons in the parton shower, it can radiate U-bosons if they are sufficiently light. Thus it might be worthwhile to look for $\mu^+ \mu^-$ pairs inside a jet, although we expect that contamination from backgrounds and difficulty obtaining good resolution on the track momenta will make this a poor competitor to the searches at high-luminosity GeV-scale experiments like BaBar, Belle, and KLOE.

The Large Hadron Collider is more useful due to its high intensity, ${\cal L} = 10^{33} - 10^{34}$cm$^{-2}$s$^{-1}$, and  the large parton density at smaller $x$. Therefore, a larger production rate, $\sim \epsilon^2 \times $prompt photon rate, is expected. Detailed studies with realistic background and detector simulations have yet to be carried out. We expect that it should be possible to reach $\epsilon \sim 10^{-3}$.  For details, see \cite{models}.

In comparison, several other medium to high energy colliders do not have the necessary luminosities to produce the U-boson under our consideration. They include: SPS, $p \bar{p}$ with $E_{\rm cm} = 630$ GeV and ${\cal L} = 2\times 10^{30}$cm$^{-2}$s$^{-1}$; ISR, $pp$ with $E_{\rm cm} = 64$ GeV and ${\cal L} = 10^{31}$cm$^{-2}$s$^{-1}$; SLC, $E_{\rm cm} \simeq 100$ GeV, ${\cal L} = 2.5 \times 10^{30} $ cm$^{-2}$s$^{-1}$; and HERA, $e$ (30 GeV) on $p$ (920 GeV), ${\cal L} = 7.5 \times 10^{31}$ cm$^{-2}$s$^{-1}$.

\subsection{Scattering with Gamma Rays}
\label{sec:gammaray}
\setcounter{footnote}{0}
Another option is to consider the process $e^- \gamma \to e^- U$. However, it is easy to see that current facilities do not offer a reasonable chance to probe this channel. Because we want a center of mass energy on the order of hundreds of MeV or a few GeV, light sources that supply hard X-rays are insufficient; one would need a gamma ray source. Gamma rays are produced by Compton backscattering of relativistic electrons from lasers in the LEGS facility at Brookhaven \cite{LEGS} and the HI$\gamma$S facility at Duke \cite{HIGS}, at rates on the order of 10$^6$ to 10$^8$ photons per second collimated in spots of about 1 cm. Such beams are insufficient for our purposes.

\subsection{Other low energy probes}

For completeness, we briefly comment on other possible low energy probes. 

QED precision measurements provides useful low energy probes of the U-boson. 
The correction due to the exchange of the U-boson is controlled by the ratio of the range of the U-boson mediated force and the size of system used for measurement, $r_{\rm QED~SYS.}$. It is typically on  the order of \footnote{We are particularly grateful to Bob Jaffe for illuminating discussions on precision QED tests, especially regarding the limits from muonic hydrogen.}
\beq
\left( \frac{m_{U}^{-1}}{r_{\rm QED~SYS.}} \right)^{p} \times \epsilon^2 e^2,
\eeq
where power $p$ depends on the system. For example, for the transitions between lower lying states of hydrogen, $r_{\rm QED~SYS.} \sim  \alpha \times 10$ eV$^{-1}$ and $p \simeq 2$ \footnote{The presence of the U-boson introduces a perturbation of $\alpha e^{- m_{\rm U} r}/r$. The shift in ground state energy of the hydrogen atom is about $\delta E/E \sim (a_B \mu)^{-2} $, where $\mu = m_{\rm U} + 2/a_B$ and $a_B$ is the Bohr radius.  Therefore, $p=2$.  Higher energy levels, in particular those with higher angular momentum,  will typically receive a correction with $p \geq 2$. 	}. Therefore, although such transitions have been measured to a very accurate level \cite{Fischer:2004jt,nist}, they still at most restrict $\epsilon \sim 10^{-1}$.  Therefore, it is useful to consider systems with much smaller size. 
The most useful among them is the measurement of $g_{\mu}-2$ \cite{gmu-2}. Based on that, Ref.~\cite{PospelovSecluded} found an upper bound $\epsilon^2 < 2 \times 10^{-5} \ (m_U/100\text{ MeV})^2$; see also \cite{gninenko-g2}.   An experiment under way using muonic hydrogen \cite{mup-exp} can in principle be sensitive to $\epsilon$, although eventually the reach will be limited by theoretical uncertainties \cite{mup-th}  to somewhat better than about $\epsilon \leq 10^{-2}$.  Currently, however, the error in theoretical prediction is dominated by proton charge radius \footnote{In fact, such experiments are designed give precise measurement of proton charge radius. } which leads to a weaker bound at the level of $\epsilon \sim 10^{-1}$.  Ref. \cite{PospelovSecluded} also found the low energy electron proton scattering, which is also sensitive to the charge radius of proton, leads to $\epsilon^2 < 4 \times 10^{-3} (m_{U}/100 \mbox{\ MeV})^2$.

Low $q^2$ elastic $\nu e$ scattering \cite{nue} can be used to set bounds on low energy U-bosons as discussed by Ref. \cite{fayet}, $g^U_{\nu} g^U_{e}/m_U^2  < G_{\rm F}$. However, in our case, the U-boson coupling to the weak neutral current is suppressed by an additional factor of $(m_U / m_Z)^2$. Therefore, the constraint will be $\epsilon^2 e^2 /m_{Z}^2 < G_{\rm F}$, which implies at most $\epsilon \leq 1$. 

Atomic parity measurements can  constrain the properties of the U-boson\cite{ap_contraints,fayet}. It needs both axial and vector coupling, $g^U_{eA} g^U_{qV} < 10^{-3} G_{\rm F}$. In our case, the axial coupling can only come from mixing with the Z and is again suppressed by a factor of $(m_U / m_Z)^2$. Therefore, while maybe slightly stronger than $\nu e$ scattering, the bound from atomic parity measurements is about $\epsilon < 10^{-1}$.

\section{Conclusion}

In this paper, we have investigated the present constraints and possibility of further experimental searches for a very weakly coupled $U(1)_d$ gauge boson, with mass $m_U \sim 100$s~MeV to 1 GeV.  We mainly focused on a set of low energy experiments, including low energy high luminosity lepton colliders, rare meson decay channels, and fixed target experiments. We found that while previous experiments lack the necessary luminosity to put useful limits on the coupling in the range of $10^{-3} -10^{-4}$, a set of current experiments should have the ability to probe the upper part of this parameter region. In particular, we urge BaBar and Belle to search for narrow peaks in $e^+ e^- \to \ell^+ \ell^- \gamma$ and $e^+ e^- \to \pi^+ \pi^- \gamma$ (near the $\rho$), KLOE to search in $\phi$ decays for narrow $\mu^+ \mu^-$ resonances, and BES-III to search in $J/\psi$ decays.

The smallest coupling $\epsilon$ that is probed scales as ${\cal L}^{-1/4}$. We argue that the best way to achieve a much higher luminosity is to design a specific low energy fixed target experiment to probe U-bosons with mass above $2m_\mu$. We discuss desired properties of such an experiment. Specifically, it is beneficial to use a thinner target, about 0.1 radiation length. The optimal reach in this case is about $10^{-6}$. At the same time, an important limitation can be the maximal rate at which relevant information of the event can be recorded, $R_{\rm max}$. The $\epsilon$ reach scales as $R_{\rm max}^{-1/4}$.  At a rate of $10^5$ Hz, which is possible, the reach on $\epsilon$ is reduced to $10^{-5}$. However, searches for long-lived U-bosons decaying outside the target can probe a range of $\epsilon$ at or below $10^{-5}$, depending on $m_U$ and experimental resolutions. We are optimistic that in this way dedicated fixed-target experiments can rule out or a discover a U-boson over nearly the entire interesting range.

\vspace{1cm}

\noindent {\bf Acknowledgements}

First and foremost, we thank N.~Arkani-Hamed for key discussions, collaboration on portions of this work, and encouragement. We thank P.~Schuster and S.~Thomas for comments on the manuscript. We are grateful to R.~Jaffe for illuminating discussions on QED precision tests, especially on muonic hydrogen. We thank T.~LeCompte and H.~Lubatti for comments on the feasibility of fixed-target experiments. We also thank T.~Han, B.~McElrath, P.~Meade, S.~Nussinov, J.~Olsen, P.~Onyisi, J.~Pivarski, M.~Pospelov, A.~Ritz, D.~Stuart, N.~Toro, and B.~Zhou for very useful discussions.  L.-T. W. is supported by the National
Science Foundation under grant PHY-0756966 and the Department of
Energy under grant DE-FG02-90ER40542.

\appendix
\section{Resolution of muon momentum measurement.}
\label{app:resolution}
\begin{figure}[!h]
\begin{center}
\includegraphics[width=8cm]{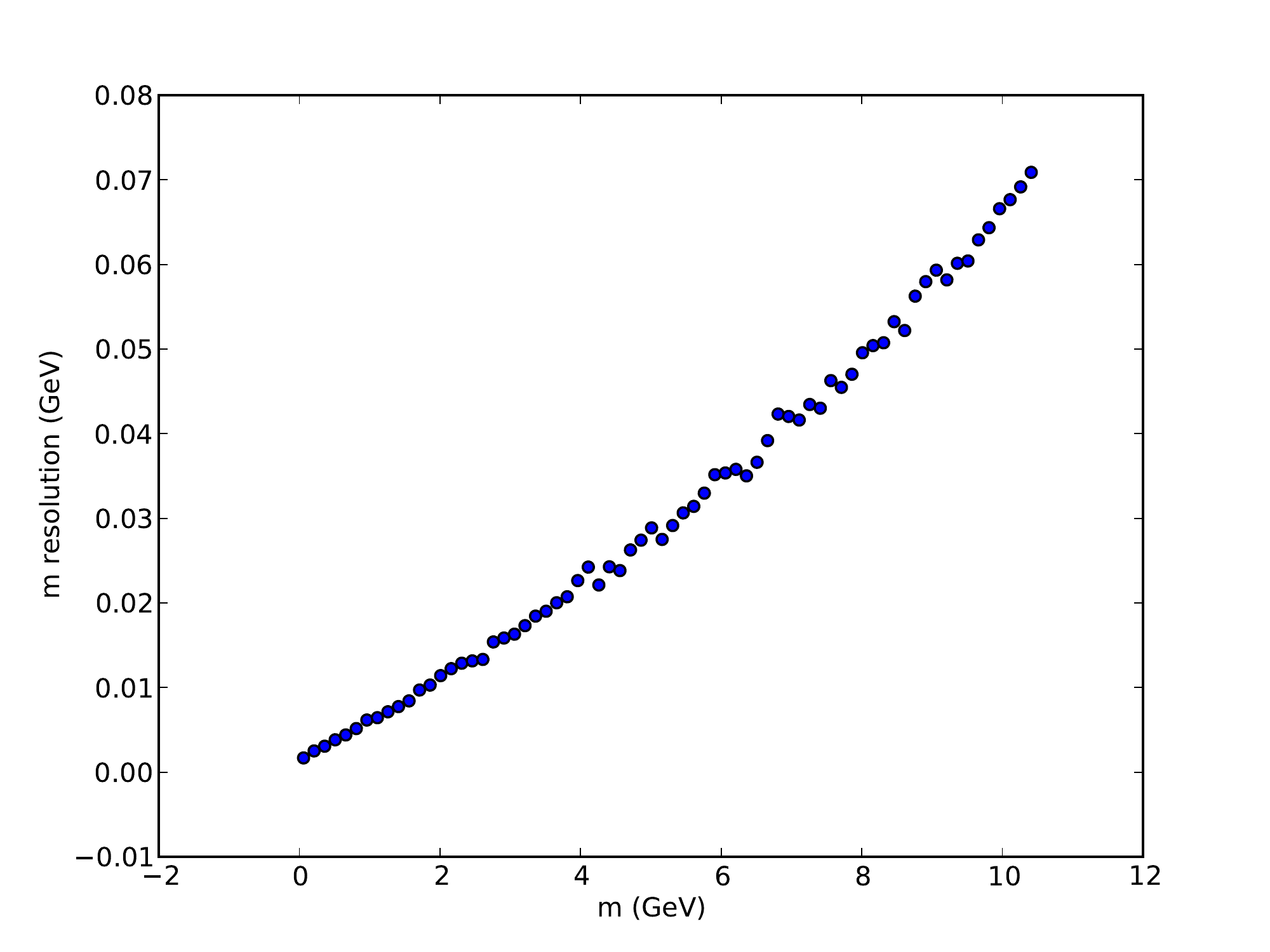}\includegraphics[width=8cm]{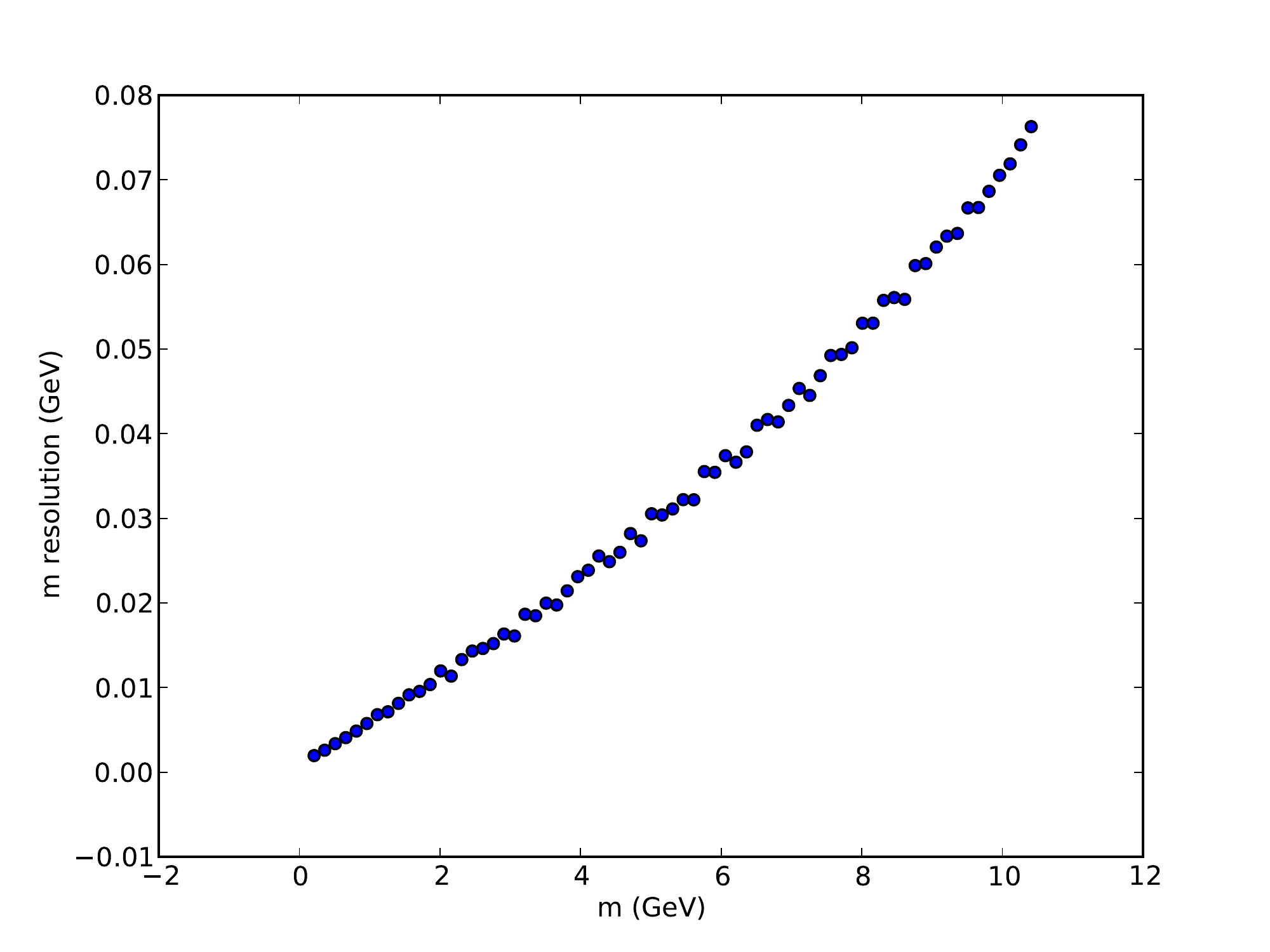}
\end{center} \caption{Resolution in invariant mass at BaBar, as a function of $m(e^+ e^-)$ at left and of $m(\mu^+ \mu^-)$ at right.} \label{fig:babarres}
\end{figure}

In the $\ell^+ \ell^- \gamma$ background samples, we have smeared the track parameters with Gaussian errors that approximate the detector reconstruction:
\beq
\sigma_{p_t}/p_t & = & 0.13\% \times p_t + 0.45\%, \\
\sigma_\phi & = & 1~{\rm mrad}, \\
\sigma_{\tan \lambda} & = & 0.001,
\eeq
where $\phi$ is the azimuthal angle and $\tan \lambda = p_z / p_t$ defines the ``dip angle" from the transverse plane. At this point the only cuts imposed on the leptons and photon are basic acceptance cuts on $\cos \theta$ (the tracks must be within the fiducial volume of the drift chamber, and the photon must hit the central electromagnetic calorimeter), $p_t$ for tracks (at least 60 MeV), and $E$ for the photon (at least 20 MeV). We assume there is no significant inefficiency for detecting tracks and photons passing these cuts, and no significant fake rates. With these assumptions, the root-mean-square mass resolution $\delta m$ is plotted against $m(\ell^+ \ell^-)$ in Figure \ref{fig:babarres}. Fitting the mass resolution curve to a quadratic polynomial, we obtain:
\beq
\delta m(e^+e^-) & = & \left(2.0 + 3.9 \left(\frac{m_U}{1.0~{\rm GeV}}\right) + 0.25 \left(\frac{m_U}{1.0~{\rm GeV}}\right)^2 \right) {\rm MeV} \\
\delta m(\mu^+\mu^-) & = & \left(1.8 + 4.1 \left(\frac{m_U}{1.0~{\rm GeV}}\right) + 0.28 \left(\frac{m_U}{1.0~{\rm GeV}}\right)^2 \right) {\rm MeV}
\eeq
 Note that to good approximation $\delta m$ is a linear function of $m_U$ up to a few GeV. As a cross-check, BaBar claims that on the $J/\psi$ peak the $\mu^+ \mu^-$ mass resolution is 14.5 MeV \cite{BaBarJPsi}. Our estimate at the $J/\psi$ mass is about 17 MeV, suggesting that our estimated resolution is approximately correct and that we err on the conservative side. The difference in resolution for electrons and muons is small but persists as we simulate larger event samples, so it appears to be due to the way the muon mass enters the invariant mass reconstructed from the tracks.
 We performed a similar fit for the $\pi^+ \pi^-$ channel, restricted to a range near the $\rho$: 590 MeV $\leq m(\pi^+ \pi^-) \leq$ 970 MeV. The resulting fit is:
 \beq
 \delta m(\pi^+ \pi^-) &=& \left(0.65 + 5.3 \left(\frac{m_U}{1.0~{\rm GeV}}\right)\right){\rm MeV}.
\eeq

\end{document}